\documentclass[letter]{aa}

\usepackage{natbib}
\usepackage{graphicx}
\usepackage{txfonts}

\begin{document} 
\bibliographystyle{aa}

   \title{Similar origin for low- and high-albedo Jovian Trojans and Hilda asteroids ?\thanks{Reflectance spectra presented in this paper are available in electronic form at the CDS via anonymous ftp to cdsarc.u-strasbg.fr (130.79.128.5) or via http://cdsweb.u-strasbg.fr/cgi-bin/qcat?J/A+A/}}
   
   \author{M. Marsset\inst{1,2}
          \and
          P. Vernazza\inst{2}
          \and
          F. Gourgeot\inst{1,3}
          \and
          C. Dumas\inst{1}
          \and
          M. Birlan\inst{4}
	  \and
	  P. Lamy\inst{2}
	  \and
	  R. P. Binzel\inst{5}
          }

   \institute{European Southern Observatory (ESO),
              Alonso de C\'ordova 3107, 1900 Casilla Vitacura, Santiago, Chile\\
              \email{mmarsset@eso.org}
         \and
             Aix Marseille University, CNRS, LAM (Laboratoire d'Astrophysique de Marseille) UMR 7326, 13388, Marseille, France
         \and
        	       	    LESIA, Observatoire de Paris, CNRS, UPMC Univ. Paris 06, Univ. Paris Diderot, 5 Place J. Janssen, 92195 Meudon, France
         \and
     IMCCE, Observatoire de Paris, 77 avenue Denfert-Rochereau, 75014 Paris Cedex, France
	  \and
	Department of Earth, Atmospheric, and Planetary Sciences, Massachusetts Institute of
Technology, Cambridge, MA 02139, USA\\
             }
             
   \date{Received XXX; accepted XXX}
 
  \abstract
   {Hilda asteroids and Jupiter Trojans are two low-albedo (${\rm p}_{\rm v}$\,$\sim$\,0.07) populations for which the Nice model predicts an origin in the primordial Kuiper Belt region. However, recent surveys by WISE and the Spitzer Space Telescope (SST) have revealed that $\sim$2\% of these objects possess high albedos (${\rm p}_{\rm v}$\,$\geq$\,0.15), which might indicate interlopers - that is, objects not formed in the Kuiper Belt - among these two populations. Here, we report spectroscopic observations in the visible and\,/\,or near-infrared spectral ranges of twelve high-albedo (${\rm p}_{\rm v}$\,$>$\,0.15) Hilda asteroids and Jupiter Trojans. These twelve objects have spectral properties similar to those of the low-albedo population, which suggests a similar composition and hence a similar origin for low- and high-albedo Hilda asteroids and Jupiter Trojans. We therefore propose that most high albedos probably result from statistical bias or uncertainties that affect the WISE and SST measurements. However, some of the high albedos may be true and the outcome of some collision-induced resurfacing by a brighter material that could include water ice. Future work should attempt to investigate the nature of this supposedly bright material. The lack of interlopers in our sample allows us to set an upper limit of 0.4\% at a confidence level of 99.7\% on the abundance of interlopers with unexpected taxonomic classes (e.g., A-, S-, V-type asteroids) among these two populations.}


   \maketitle
%

\section{Introduction}
\label{sec:introduction}

Jupiter Trojans and Hilda asteroids are small primitive bodies located beyond the snow line, around the L4 and L5 Lagrange points of Jupiter at $\sim$\,5.2 AU (Trojans) and in the 2:3 mean-motion resonance with Jupiter near 3.9\,AU (Hilda asteroids). Their origin remains a major challenge to current theories of the formation of the solar system.

There are two current models, the Grand Tack (\citealt{Walsh:2011co}) and the Nice model (\citealt{Morbidelli:2005dr, Levison:2009jta}), each one addressing distinct epochs of the early dynamical evolution of the solar system, and which - when taken together - make key predictions on the origin of both the Hilda and the Jupiter Trojan populations. The Grand Tack model addresses the early dynamical evolution of the solar system 3 to 5 Myr after its formation; it suggests that Jupiter roamed inward as close as the present location of Mars ($\sim$1.5\,AU) and then outward. This migration profoundly influenced the solar system, causing a substantial radial mixing of planetesimals throughout the solar system and making Mars smaller than it should have been. The Nice model addresses the late dynamical evolution of the solar system $\sim$700 Myr after its formation; it suggests that a large portion of both the Trojans and the Hilda asteroids formed in more distant regions - typically in the primordial transneptunian disk, the precursor of the Kuiper Belt - and subsequently chaotically migrated towards the inner solar system as a consequence of the outward migration of all four giant planets. 

Taken together, these models suggest that the immediate precursors of both the Hilda asteroids and Jupiter Trojans are Kuiper Belt objects (Nice model) and that the latter are not a homogeneous population (Grand Tack model), comprising a large portion of objects formed at large heliocentric distances ($\geq$\,10\,AU) and a minor fraction of planetesimals formed closer to the Sun ($\leq$\,3\,AU). The Trojans and Hilda asteroids could therefore represent a condensed or mixed version of the primordial solar system, very much like the asteroid belt. 

Until recently, telescopic observations of both populations had only focused on the brightest, that is, the largest, objects. Their spectroscopic observations in the visible \citep{Dahlgren:1995tp, Dahlgren:1997uta, Bus:2002bea, Lazzaro:2004ja, Fornasier:2004if, Fornasier:2007kf, Dotto:2006kx, Roig:2008jx} and near-infrared \citep{Dumas:1998jn, Dotto:2006kx, Yang:2011fl, Emery:2011kr} have revealed uniformly red and featureless spectra, and their albedo measurements \citep{Fernandez:2003kl} have indicated low values ($\sim$0.07) for both populations. 

However, more recent measurements with the Spitzer Space Telescope (SST, \citealt{Fernandez:2009joa}) and with WISE \citep{Grav:2011eqa, Grav:2012ho, Grav:2012cl} have revealed a small group ($\sim$1.7$\%$ of the total population) of presumably bright (${\rm p}_{\rm v}$\,$\geq$\,0.15) objects among both Hilda asteroids and Jupiter Trojans, which might indicate a minor fraction of interlopers - that is, objects not originating from the Kuiper Belt region, with a reflectance spectrum different from the typical X\,/\,T\,/\,D-class asteroids - among these two populations.  

In this paper, we report the very first spectroscopic characterization of a sample of high-albedo Hilda asteroids and Trojans.

\section{Observations and data reduction}
\label{sec:observations}

The spectra presented in this work were collected between April and December 2013 using two different instruments, X-SHOOTER (ESO/VLT, Chile) and SpeX (NASA/IRTF, Hawaii). A total of 17 spectra for 12-high-albedo Hilda asteroids and Trojans (with geometric albedo ${\rm p}_{\rm v}$\,$\ge$\,0.15) were obtained; in particular, a few objects were observed twice with both instruments. All observations were performed under clear sky conditions with the seeing being predominantly in the $0.6''$ to $1.2''$ range. Observation circumstances are summarized in Table \ref{tab:obs}. 

\subsection{VLT/X-SHOOTER}
\label{sec:xsh}

Observations with X-SHOOTER were performed in visitor mode. X-SHOOTER is a medium-resolution {\it Echelle} spectrograph that covers the 0.3$-$2.5\,{\rm$\mu$m} spectral range in one shot by splitting the incoming light into three beams, each sent to a different arm (UVB, VIS, NIR) of the instrument \citep{Vernet:2011bya}. We used the $1.6''$, $1.5''$ and $1.2''$ slits for the UVB, VIS and NIR arms respectively, each arm has a resolving power in the 3000$-$5500 range. To monitor the high luminosity and variability of the sky in the near-IR, the telescope was moved along the slit during the acquisition of the data so as to obtain a sequence of spectra located at two different positions on the array. These paired observations provided near-simultaneous sky and detector bias measurements. 

The asteroid observations were interspaced with those of the standard stars SA\,102-1081, SA\,98-978, SA\,93-101, and Hyades\,64 \citep{Landolt:1973br}. These solar analogs were observed at similar airmass with respect to our asteroid targets to ensure a good correction for atmospheric extinction. 

Note that the slit was aligned with the parallactic angle for all asteroid and star observations to limit the loss of flux at short wavelengths due to differential atmospheric refraction.

Data reduction - which included bad-pixel mapping and correction, flat-fielding, wavelength calibration, merging of the different orders, and subtraction of dithered pairs - was performed with the X-SHOOTER ESO pipeline (v2.3.0) using the ESO reflex environment \citep{Freudling:2013cp}. After they were extracted, the asteroid spectra were divided by those of the solar analogues to produce reflectance spectra that were smoothed with a median-filter technique, using a box of 7 pixels in the spectral direction for each point of the spectrum.

Note that the near-IR data were obtained using the non-destructive readout mode of the detector, in which each independent pixel has a threshold limit of about 42\,kADUs that - if reached before the end of the integration time - causes the ADU level to be extrapolated from values below the threshold. This mode avoids saturation but can return an incorrect extrapolation of the signal, especially if the observing conditions are variable. An analysis of our raw spectra revealed that, during some observations, the {\it K}-band signal did reach the ADU threshold limit because of the contribution of the sky background, which resulted in a potentially incorrect extrapolation (in our case a drop in flux) of $\sim$\,5 to 10\% of the signal in the 2.0$-$2.5\,$\mu$m range. The affected spectra are flagged in Table \ref{tab:obs} to prevent alteration of our final analysis (see Section \ref{sec:results}).

\subsection{IRTF/SpeX}
\label{sec:spex}

IRTF observations of some objects in our sample (Table \ref{tab:obs}) were carried out remotely from Europe. The low- to medium-resolution near-IR spectrograph and imager SpeX \citep{Rayner:2003hfa} combined with a 0.8\,$\times$\,15 arcsec slit was used in the low-resolution prism mode to acquire the spectra in the 0.7$-$2.5 \,{\rm$\mu$m} wavelength range. Our solar analogue stars were SA\,93-101, Hyades\,64, and SA\,98-978 \citep{Landolt:1973br}. Standard techniques for near-IR spectroscopy data reduction (see above) were used with the software Spextool \citep{Cushing:2004bq} to extract the asteroid and solar analogue spectra. Finally, the asteroid reflectance was obtained by dividing each asteroid average spectrum by the average solar star spectrum.

\begin{table*}[tbh]
\scriptsize
\begin{center}
\caption[]{\em{\small Observation circumstances}}
\label{tab:obs}
\begin{tabular}{llccccccccc}
\hline \noalign {\smallskip}
No. & Name & Group & Date & Instrument & AM & Exp. time (s) & {\it D} (km)  & ${\rm p}_{\rm v}$ & Ref. & \\
\hline \noalign {\smallskip}
1162 & 1930 AC                   & Hilda   & 2013-08-29 & X-SHOOTER & 1.464$-$1.501 & 2$\times$880/900/900 & 40.38 $\pm$ 0.30 & 0.186 $\pm$ 0.037 & 1 & \\
	&		                    &	         & 2013-09-01 & SpeX		& 1.038$-$1.074 & 10$\times$120			& 				 &				  & & \\
3290 & 1973 SZ$_1$           & Hilda   & 2013-08-28 & X-SHOOTER & 1.201$-$1.277 & 3$\times$880/900/900 & 10.18 $\pm$ 0.30 & 0.324 $\pm$ 0.055 & 1 & \\
	&				   &		 & 2013-08-29 & X-SHOOTER & 1.316$-$1.422 & 880/900/900			&			       &				& &  \\
5023 & 1985 TG$_3$           & Trojan & 2013-04-13 & X-SHOOTER & 1.400$-$1.716 & 3$\times$880/900/900 & 27.85 $\pm$ 3.51 & 0.173 $\pm$ 0.093 & 2 & * \\
9713 & 1973 SP$_1$           & Trojan & 2013-04-14 & X-SHOOTER & 1.370$-$1.652 & 3$\times$880/900/900 & 18.65 $\pm$ 3.38 & 0.168 $\pm$ 0.086 & 2 & * \\
11249 & 1971 FD                 & Hilda   & 2013-08-28 & X-SHOOTER & 1.065$-$1.070 & 2$\times$880/900/900 & 9.97 $\pm$ 0.88 & 0.371 $\pm$ 0.097 & 1 & \\
	  &				   &		 & 2013-08-29 & X-SHOOTER & 1.067$-$1.077 & 880/900/900			&			       &				& &  \\
	  &		                    &            & 2013-09-01 & SpeX               & 1.288$-$1.631 & 26$\times$120			&                            &                                & & \\
13331 & 1998 SU$_{52}$    & Trojan & 2013-04-15 & X-SHOOTER & 1.313$-$1.460 & 3$\times$880/900/900 & 17.68 $\pm$ 1.54 & 0.171 $\pm$ 0.033 & 2 & * \\
14669 & 1999 DC                & Hilda   & 2013-12-13 & SpeX 		& 1.002$-$1.023 & 30$\times$120			& 16.11 $\pm$ 0.66 & 0.206 $\pm$ 0.038 & 1 & \\
15529 & 2000 AA$_{80}$    & Trojan & 2013-04-16 & X-SHOOTER & 1.295$-$1.431 & 3$\times$880/900/900 & 16.43$\pm$ 1.33 & 0.198 $\pm$ 0.093 & 2 & * \\
24452 & 2000 QU$_{167}$ & Trojan & 2013-08-28 & X-SHOOTER & 1.402$-$1.426 & 880/900/900 & 18.69 $\pm$ 0.99 & 0.184 $\pm$ 0.029 & 2 & \\
	  &				   &		& 2013-08-29 & X-SHOOTER & 1.326$-$1.333 & 880/900/900		&			       &				& & \\
32430 & 2000 RQ$_{83}$   & Trojan & 2013-08-29 & X-SHOOTER & 1.447$-$1.520 & 2$\times$880/900/900 & 13.37 $\pm$ 0.55 & 0.157 $\pm$ 0.007 & 2 & \\
63284 & 2001 DM$_{46}$   & Trojan & 2013-04-13 & X-SHOOTER & 1.267$-$1.548 & 3$\times$800/900/900 & 12.27 $\pm$ 0.68\,/\,10.26 $\pm$ 1.32 & 0.129 $\pm$ 0.026\,/\,0.252 $\pm$ 0.050 & 1\,/\,3 & * \\
	  &				   &		& 2013-04-15 & X-SHOOTER & 1.275$-$1.593 &  3$\times$880/900/900		&			       &				& & * \\
65227 & 2002 ES$_{46}$    & Trojan & 2013-04-14 & X-SHOOTER & 1.335$-$1.626 & 3$\times$880/900/900 & 13.58 $\pm$ 1.51\,/\,14.04 $\pm$ 0.10 & 0.126 $\pm$ 0.035\,/\,0.179 $\pm$ 0.003 & 1\,/\,3 & * \\
	  &				   &		& 2013-04-16 & X-SHOOTER & 1.341$-$1.487 &  3$\times$880/900/900		&			       &				& & * \\

\hline \noalign {\smallskip}
\end{tabular}
\begin{flushleft}
{\bf Notes.}  {\it D}: effective diameter; ${\rm p}_{\rm v}$: optical albedo. Exp. times for X-SHOOTER are given for the UVB/ VIS/NIR arms. The asterisk indicates objects for which the signal has been partially extrapolated in the K-band. \\ 
{\bf References.} 1: \citealt{Grav:2012ho}; 2: \citealt{Grav:2011eqa, Grav:2012cl}; 3: \citealt{Fernandez:2009joa}. \\
\end{flushleft}
\end{center} 
\end{table*}

\section{Results}
\label{sec:results}

\begin{figure*}
\centering
\includegraphics[angle=0,width=0.8\linewidth, trim=0cm 8cm 0cm 1.6cm, clip]{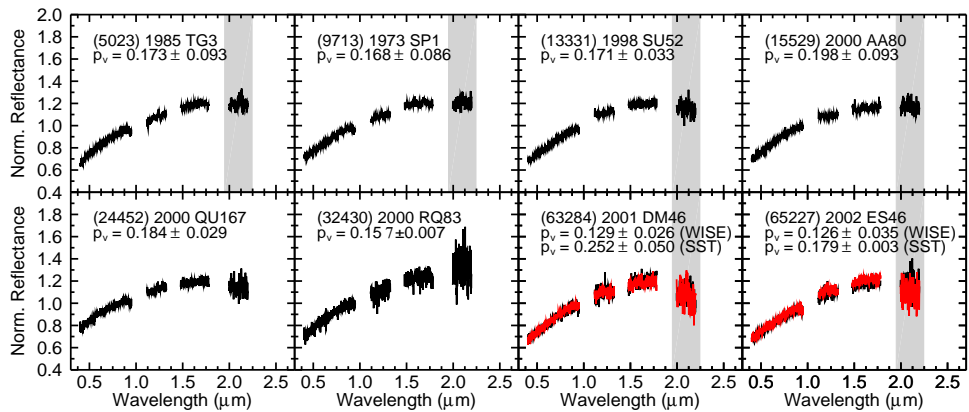}
   \caption{X-SHOOTER spectra of Jupiter Trojans normalized to unity at 1\,$\mu$m. Second spectrum from multiple observations of the same asteroid are overplotted in red. The grey bands indicate possible saturation in the X-SHOOTER data (see Section\,\ref{sec:xsh}).
           }
      \label{fig:trojans}
\end{figure*}

\begin{figure*}
\centering
\includegraphics[angle=0,width=0.8\linewidth, trim=0cm 11.5cm 0cm 1.6cm, clip]{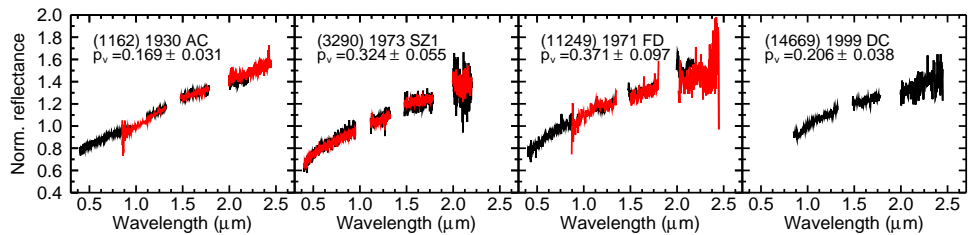}
   \caption{X-SHOOTER and SpeX spectra of Hilda asteroids normalized to unity at 1\,$\mu$m.
           }
      \label{fig:hildas}
\end{figure*}

\begin{figure}[ht]
\centering
\includegraphics[trim=0mm 2mm 0mm 5mm, clip, width=0.43\textwidth]{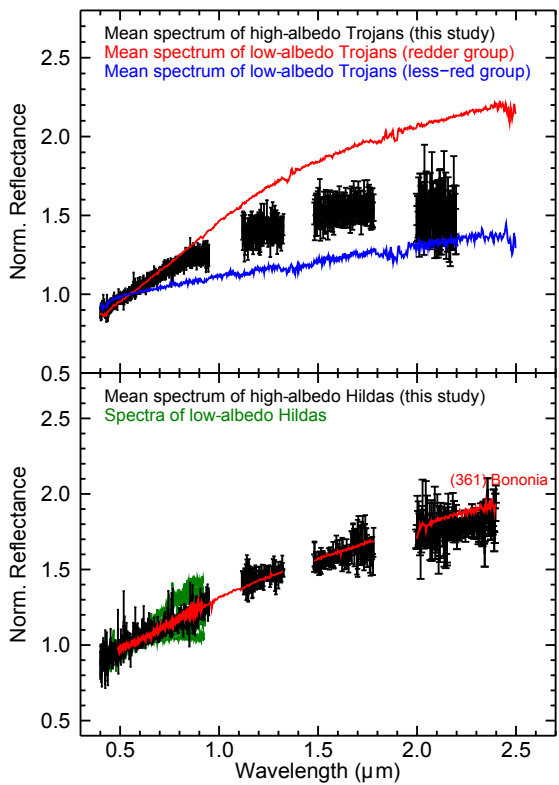}
\caption{All spectra are normalized to unity at 0.55\,$\mu$m. (Top) Mean spectra of the two groups identified by \citet{Emery:2011kr} among low-albedo Trojans compared with the mean spectrum of bright Trojans (from this study). (Bottom) Visible spectra of Hilda asteroids from \citet{Bus:2002bea}, \citet{Lazzaro:2004ja}, and the VNIR spectrum of (361) Bononia \citep{Bus:2002bea, Takir:2012cza} compared with the mean spectrum of bright Hilda asteroids (from this study).} 
\label{fig:mean}
\end{figure}

The X-SHOOTER and SpeX reflectance spectra were normalized to unity at 1\,{\rm$\mu$m} and are shown in Fig.\,\ref{fig:trojans} for Trojans and in Fig.\,\ref{fig:hildas} for Hilda asteroids. A comparison of our spectra with those of low-albedo Hilda asteroids and Jovian Trojans \citep{Dahlgren:1995tp, Dahlgren:1997uta, Fornasier:2004if, Fornasier:2007kf, Roig:2008jx, Emery:2011kr} suggests that both populations have very similar spectral properties (Figure\,\ref{fig:mean}). This spectral similarity is confirmed by placing the objects into the taxonomic system of \citet{Bus:1999tw} for the X-SHOOTER spectra (using only the visible portion of our spectra to avoid the saturation problem in the K band) and in the \citet{DeMeo:2009gz} system for the near-IR (SpeX) spectra. Whereas all Trojans appear to be D-types based on the visible range alone, a comparison of their complete visible and near-IR spectra with the average \citet{DeMeo:2009gz} spectra for X-, T- and D-types reveals that their average spectrum is closer to the T-type. The only exception is (24452) 2000 QU$_{167}$, which has a bluer slope that is consistent with X-types. Hilda asteroids have slightly redder slopes than Trojans, with values in between those of the T- and D-type spectra.\footnote{Note that most authors preferentially use a broader definition of the X- and D- classes instead of using the T-class (e.g., \citealt{GilHutton:2008de, DeMeo:2014hz}), which explains the scarcity of T-type classification of asteroids in the literature.}

A careful inspection of the spectra furthermore reveals that none of them displays any apparent absorption feature within the level of noise of our data set. The absence of absorption bands in the spectra of Trojans was also reported by \citet{Emery:2011kr} for low-albedo objects and is consistent with the few spectra that have been acquired for low-albedo Hilda asteroids \citep{Dahlgren:1995tp, Dahlgren:1997uta, Dumas:1998jn, Bus:2002bea, Takir:2012cza}.

\begin{figure}
\centering
\includegraphics[trim=0mm 0mm 0mm 15mm, clip, angle=0,width=0.8\linewidth]{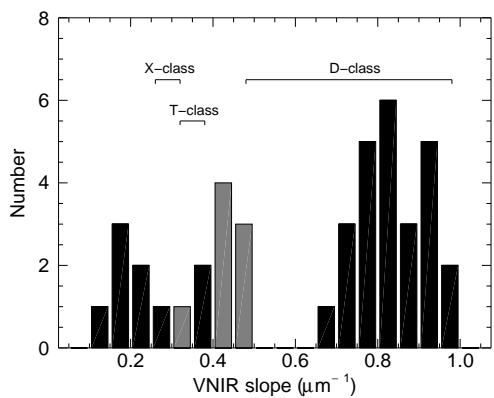}
   \caption{Histogram of VNIR (0.55-1.80\,$\mu$m) spectral slopes for Trojans. Black corresponds to data from previous studies and grey to new data from this paper. The {\bf 1\,$\sigma$} bounds of the X-, T- and D- classes from \citet{DeMeo:2009gz} are shown.
           }
      \label{fig:slopes}
\end{figure}

As a last step, we performed a more detailed comparison of our dataset with previously published spectra of low-albedo Trojans (this is not possible for Hilda asteroids as both visible and near-IR spectra have not yet been collected for a statistically significant sample). Specifically, we calculated the spectral slope over the broadest possible wavelength range (0.55$-$1.8\,${\rm \mu}$m) both for our Trojan sample and for the low-albedo objects measured by \citet{Emery:2011kr}. The near-IR spectra of these authors were complemented by the visible spectra of \citet{Vilas:1993ij}, \citet{Xu:1995gy}, \citet{Bus:2002bea}, \citet{Lazzaro:2004ja}, \citet{Fornasier:2004if}, and \citet{Fornasier:2007kf}. Figure\,\ref{fig:slopes} shows that our spectra fall within the range of the blue Trojans, although they are on average redder. Note that they are not sufficiently red to fill the gap between the blue and the red Trojans that results from the bimodality reported by \citet{Emery:2011kr}, yet they narrow the gap to some extent. The redder colours of our sample with respect to the blue Trojans from \citet{Emery:2011kr} are essentially due to steeper slopes in the UVB/VIS range.

\section{Discussion}
\label{sec:discussion}
Our study showed that the observed twelve high-albedo Hilda asteroids and Jupiter Trojans have the same spectral properties in the visible and near-IR ranges as the low-albedo objects. The lack of spectral differences in our sample may suggest that the high albedos were incorrectly derived by SST and WISE. This would not be surprising considering the natural broadening of the Gaussian error with the increasing number of objects at smaller sizes (for a more detailed discussion, see \citealt{Grav:2011eqa, Grav:2012ho}). The few high-albedo objects would therefore correspond to the outlier measurements for which the statistical error pulled towards higher signal.

Alternatively, if the albedo measurements are valid, we propose as a first suggestion that they result from a collision-induced resurfacing process that exposed bright material at the surface. Because an origin in the primordial transneptunian disk is predicted for these objects \citep{Morbidelli:2005dr, Levison:2009jta}, this bright material could include water ice. However, our spectra do not show any evidence for this material at the surface of these objects. More studies in the 3\,$\mu$m region may shed light on the nature of this putative bright material.

The lack of spectral differences in our sample allows setting an upper limit on the fraction of potential interlopers within Jupiter Trojan and Hilda populations (i.e., inner solar system small bodies such as S-type asteroids) and thus constraints on migration models. If the high albedos result from statistical measurement errors, both our study and the WISE survey indicate that there are no interlopers among Trojans and Hilda asteroids. Conversely, if the high albedos are valid, we find that among the fifty bright Jupiter Trojans and Hilda asteroids detected by SST and WISE, and considering that twelve of them are X-, T- or D-types, at most $\sim$\,22\% of them can be interlopers at the 99.7\% confidence level (this result was derived from the hypergeometric probability distribution). Therefore, at most $\sim$0.4\% of the $\sim$2900 objects surveyed by SST and WISE may be interlopers at the 99.7\% confidence level. The paucity or absence of interlopers among Jupiter Trojans and Hilda asteroids revealed by our study is consistent with the results of \citet{DeMeo:2013hw}, who found no evidence of S-types or other unexpected classes (e.g., A-, V-) among them based on the SDSS survey.

\section{Conclusions}
\label{sec:conclusions}

We have measured the visible and\,/\,or near-IR spectral properties for twelve high-albedo (${\rm p}_{\rm v}$\,$\geq$\,0.15) Hilda asteroids and Jupiter Trojans that are part of the fifty high-albedo objects detected by SST and WISE out of a total of 2875 objects surveyed in both populations. These twelve objects have the same spectral properties as their low-albedo counterparts, which suggests a similar origin for low- and high-albedo Hilda asteroids and Jupiter Trojans. The lack of interlopers in our sample allowed us to place an upper limit of 0.4\% at a confidence level of 99.7\% on the fraction of interlopers  (e.g., A-, S-, V-type asteroids) among these two populations. This result is based on the assumption that albedos were accurately determined by SST and WISE. If this were not the case, both Hilda asteroids and Jupiter Trojans probably comprise no interlopers. Future work should attempt to determine the cause of the high albedos.\\

\begin{acknowledgements}
We are grateful to Josh Emery for providing his observations of 66 Trojans and to Driss Takir for providing his observations of (153) Hilda, (190) Ismene, and (361) Bononia. We also thank the referee, Tommy Grav, for his very valuable comments, and Francesca DeMeo for helping us clarify some ambiguities. Finally, we acknowledge the ESO VLT and NASA IRTF staff for their assistance with asteroid observations.
\end{acknowledgements}


\bibliography{references}

\end{document}